\documentclass[11pt]{article}

\usepackage[final]{acl}

\usepackage{times}
\usepackage{latexsym}
\usepackage{booktabs}
\usepackage{multirow}
\usepackage{subcaption}
\usepackage{amsmath}
\usepackage{amssymb}
\usepackage{xcolor}
\usepackage[T1]{fontenc}

\usepackage[utf8]{inputenc}

\usepackage{microtype}

\usepackage{inconsolata}

\usepackage{graphicx}

\newcommand{\ours}{\texttt{R}$^\texttt{3}$\texttt{AG}}

\title{R$^3$AG: Retriever Routing for Retrieval-Augmented Generation}

\author{
Tong Zhao\textsuperscript{1},
Yutao Zhu\textsuperscript{1}\thanks{Corresponding author.},
Yucheng Tian\textsuperscript{2},
Zhicheng Dou\textsuperscript{1} \\
\textsuperscript{1}Renmin University of China \\
\textsuperscript{2}South China University of Technology \\
\texttt{zhaotong7@ruc.edu.cn, yutaozhu94@gmail.com} \\
}

\begin{document}
\maketitle
\begin{abstract}
Retrieval-augmented generation (RAG) has become a cornerstone for knowledge-intensive tasks. However, the efficacy of RAG is often bottlenecked by the ``one-size-fits-all'' retrieval paradigm, as different queries exhibit distinct preferences for different retrievers. While recent routing techniques attempt to select the optimal retriever dynamically, they typically operate under a ``single and static capability'' assumption, selecting retrievers solely based on semantic relevance. This overlooks a critical distinction in RAG: a retrieved document must not only be relevant but also effectively support the generator in producing correct answers. To address this limitation, we propose \ours{}, a novel routing framework that explicitly models the dynamic alignment between queries and retriever capabilities. Unlike previous approaches, \ours{} decomposes retriever capability into two learnable dimensions: retrieval quality and generation utility. We employ a contrastive learning objective that leverages complementary supervision signals, \textit{i.e.}, document assessments and downstream answer correctness, to capture query-specific preference shifts. Extensive experiments on several knowledge-intensive tasks show that \ours{} consistently outperforms both the best individual retrievers and state-of-the-art static routing methods.
\end{abstract}

\section{Introduction}\label{sec:intro}
\begin{figure}[t]
    \centering
    \includegraphics[width=\linewidth]{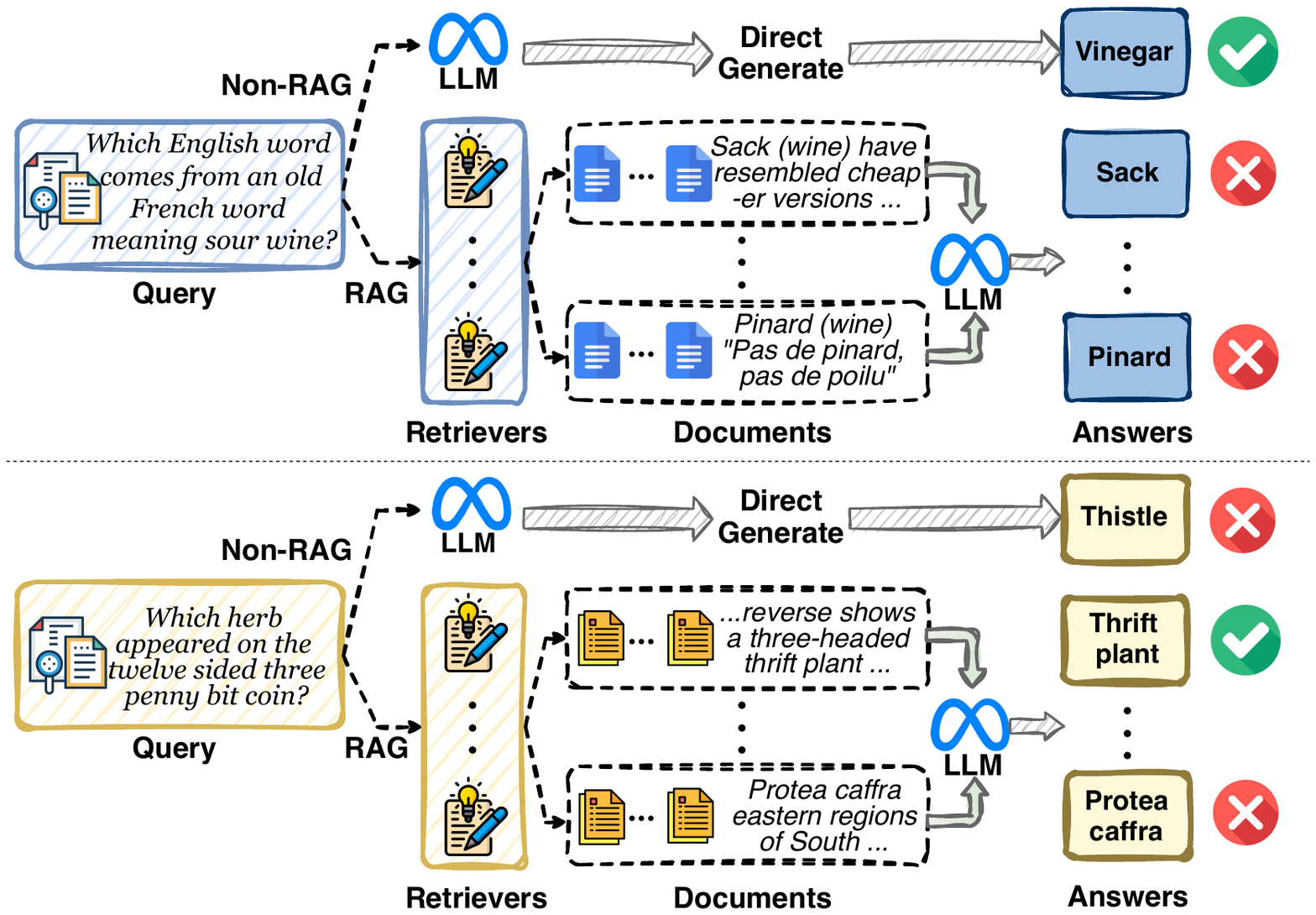}
    \caption{(Top) Retrieval is not universally beneficial. (Bottom) Different
    retrievers yield different generation results of varying quality.}
    \label{fig:intro}
\end{figure}
The rapid progress of large language models (LLMs) has substantially advanced the state-of-the-art in natural language processing(NLP), playing an increasingly important role in a wide range of real-world applications~\citep{DBLP:conf/nips/VaswaniSPUJGKP17,radford2018improving,DBLP:conf/nips/BrownMRSKDNSSAA20,DBLP:journals/corr/abs-2303-12712,DBLP:journals/corr/abs-2303-18223}. Despite these impressive capabilities, LLMs are inherently constrained by their static and finite training corpora. Consequently, they frequently suffer from critical limitations such as hallucinations and inadequate timeliness, particularly when confronted with knowledge-intensive tasks~\citep{ji2023survey,DBLP:journals/corr/abs-2301-00303,openai2024gpt4technicalreport}. In response to these challenges, retrieval-augmented generation (RAG)~\citep{lewis2020retrieval,DBLP:conf/icml/GuuLTPC20} has emerged as the \textit{de facto} gold standard. By incorporating external evidence during inference, RAG effectively bridges the gap between the static training corpus of LLMs and the dynamic, knowledge-intensive characteristics of real-world queries~\citep{cai2022recent,gao2023retrieval,DBLP:conf/eacl/IzacardG21,singh2021end,DBLP:journals/access/HindiMMA25}.


However, the efficacy of RAG is heavily contingent on its retrieval component, which remains a significant bottleneck. In practice, most systems adopt a ``one-size-fits-all'' approach, utilizing a fixed retriever regardless of the input query. This strategy is suboptimal for two primary reasons. On the one hand, retrieval is not always necessary~\citep{DBLP:conf/emnlp/PetroniRRLBWM19}: when the model's internal parametric knowledge is sufficient, retrieving additional evidence may introduce noise and degrade generation performance, as shown in Figure~\ref{fig:intro}. On the other hand, extensive research reveals that no single retriever outperforms all others across all query types~\citep{DBLP:conf/aaai/LeeSCSL25,DBLP:journals/corr/abs-2506-13743}. For example, sparse retrievers often excel at precise entity matching, whereas dense retrievers are superior at capturing semantic nuance. Therefore, the optimal retriever choice is highly query-dependent. This necessitates a move towards adaptive retrieval strategies and motivates our core research question: \textit{How can we effectively route queries to the most suitable retriever under the RAG paradigm?}

Recently, several studies have attempted to design query-oriented routers to address this challenge~\citep{peng2025learningroutequeriesknowledge,hsu2025large,DBLP:conf/aaai/LeeSCSL25,DBLP:journals/corr/abs-2506-13743}. These approaches generally model retriever capability through a single lens, such as semantic relevance, and assume a fixed performance profile for each retriever. For example, some methods simply determine whether to invoke a text or table retriever based solely on the data type required by the query~\citep{peng2025learningroutequeriesknowledge}. While promising, such methods overlook the dual nature of the RAG objective: retrieved documents must not only be relevant (i.e., achieve high retrieval quality) but also utility-driven for the generator to ensure strong generation performance. In other words, a document deemed relevant by standard IR metrics may still lead to erroneous generation if it contains conflicting information or lacks the specific reasoning pattern required by the LLM. Consequently, routing strategies that neglect the complex interplay between retrieval quality and generation utility fail to adapt to the dynamic requirements of RAG scenarios.

To address this problem, we propose \ours{}, a \underline{{R}}etriever \underline{{R}}outing method for \underline{{RAG}}. \ours{} is a contrastive learning-based framework explicitly designed to model the dynamic alignment between queries and retriever capabilities. Unlike prior methods that treat retrievers as black boxes with static scores, \ours{} decomposes retriever capability into two learnable dimensions: \textit{retrieval quality}, representing the ability to find high-quality evidence, and \textit{generation utility}, representing the ability to support accurate downstream answer generation. At the architectural level, \ours{} employs a multi-head attention mechanism to fuse these capability dimensions under the guidance of the input query, producing a hybrid representation that captures the query-specific preferences of different retrievers. For optimization, we devise a contrastive learning objective where positive and negative samples are constructed based on two complementary supervision signals: the quality of the retrieved documents and the correctness of downstream answer generation. This objective prompts the alignment between queries and retrievers that can simultaneously provide high-quality evidence and enable correct answer generation, thereby effectively guiding the router to prioritize end-to-end RAG performance over simple relevance matching.

We conduct a comprehensive suite of knowledge-intensive tasks~\citep{DBLP:conf/acl/JoshiCWZ17,DBLP:journals/tacl/KwiatkowskiPRCP19,DBLP:conf/emnlp/Yang0ZBCSM18} of \ours{}. Experimental results demonstrate that \ours{} consistently outperforms both the best individual retriever and state-of-the-art static routing methods. These results validate the superiority of dynamic routing and the critical role of generation utility in RAG.

Our main contributions are threefold:

(1) We identify the limitations of the single and static capability assumption in existing methods, highlighting the necessity of distinguishing between retrieval relevance and generation utility.

(2) We propose \ours{}, a novel routing framework that explicitly decomposes retriever capability into learnable retrieval quality and generation utility embeddings. By leveraging contrastive learning, \ours{} effectively captures dynamic and query-specific retriever preferences.

(3) Extensive experiments on knowledge-intensive tasks demonstrate that \ours{} successfully exploits retriever strengths, consistently outperforming both the best individual retrievers and state-of-the-art static routing methods.

\section{Related Work}
\textbf{RAG}\quad
RAG effectively mitigates the knowledge limitations of LLMs by incorporating external evidence at inference time~\citep{lewis2020retrieval,gao2023retrieval}. To enhance RAG performance, prior research has focused on optimizing specific components, such as refining the retriever~\citep{DBLP:conf/acl/YuXY023}, or improving the generator's utilization of context through query rewriting~\citep{ma2023query, peng2024large}, re-ranking~\citep{macavaney2020efficient}, and context compression~\citep{DBLP:conf/icml/LiZ0X00C25}. More recently, the field has witnessed a shift towards diverse retrieval capabilities, including hybrid indexing strategies~\citep{DBLP:journals/corr/abs-2509-21336} and multi-perspective retrieval paradigms~\citep{DBLP:journals/corr/abs-2502-18139}. This increasing heterogeneity in retrieval methods reveals a critical opportunity: rather than relying on a single retriever, systems can benefit from ensemble approaches that strategically leverage multiple retrievers. This motivation underscores our focus on the retriever routing problem within RAG settings.

\noindent\textbf{Routing Techniques}\quad 
Routing mechanisms have been extensively explored to improve model efficiency and performance. In the context of LLMs, routing primarily involves selecting domain-specific experts or models, utilizing either jointly trained routers~\citep{DBLP:journals/corr/abs-2403-07816,DBLP:conf/icml/MuqeethLLR24} or post-hoc techniques based on domain signals~\citep{DBLP:conf/iclr/FengSBBHT24,DBLP:conf/aiccc/Belofsky23,DBLP:conf/acl/ShenLWKS24}. In the realm of information retrieval, routing is often framed as federated search, where sub-queries are assigned to expert retrievers or domain-specific indices~\citep{DBLP:conf/emnlp/LinL0K23,DBLP:conf/aaai/LeeSCSL25}. Within RAG specifically, recent efforts have investigated adaptive strategies, such as determining whether to retrieve~\citep{DBLP:conf/acl/MallenAZDKH23, DBLP:conf/naacl/JeongBCHP24} or routing between different LLMs~\citep{DBLP:journals/corr/abs-2505-23052}. However, these approaches typically operate under a static assumption or assess retrievers solely based on retrieval relevance. They overlook a crucial factor in RAG: a retriever's value should be measured by its utility in supporting downstream generation. To bridge this gap, our proposed \ours{} explicitly models both retrieval quality and generation utility, enabling more effective query routing in RAG scenarios.

\section{Problem Formulation}
RAG improves LLMs by incorporating external knowledge via a external supplemental framework. Given a query $q$ and an external corpus $\mathcal{D}$, a retriever $R$ retrieves a set of relevant documents $d = R(\mathcal{D}, q)$ from the corpus. Subsequently, a generator $M$ produces a response $y$ conditioned on both the query and the retrieved documents, formulated as $y = M(q, d)$.

We focus on the problem of \textit{retriever routing} within the RAG framework. Let $\mathcal{R} = \{R_1, \ldots, R_K\}$ denote a pool of $K$ candidate retrievers. To account for scenarios where external knowledge is unnecessary or potentially detrimental, we introduce a special null retriever $R_0$, which returns an empty document set, \textit{i.e.}, $R_0(\mathcal{D}, q) = \varnothing$. In this case, the generation reduces to the standard parametric setting $y = M(q)$, making the non-RAG setting a special case of our framework.

The goal of \textbf{retriever routing} is to select the optimal retriever index for a given query. We define a routing policy $\pi: \mathcal{Q} \rightarrow \{0, 1, \ldots, K\}$ that maps a query $q$ to an index $i$. If $\pi(q) = i > 0$, the $i$-th retriever is invoked; if $\pi(q) = 0$, no retrieval is performed.




To evaluate the quality of a routing decision, we introduce a utility scoring function $s(y, y^*) \in [0, 1]$, which measures the quality of the generated response $y$ against a reference answer $y^*$. Unlike standard IR metrics that measure retrieval relevance, this function explicitly quantifies the \textit{generation utility} of a retriever. The objective of the routing policy is to maximize the expected generation performance across the query distribution:
\begin{equation}
\max_{\pi} \; \mathbb{E}_{q \sim \mathcal{Q}} \Big[ s\big(M(q, R_{\pi(q)}(\mathcal{D}, q)), y^*\big) \Big].
\end{equation}
This formulation implies that the optimal router must dynamically adapt to query needs, selecting a specific retriever or bypassing retrieval entirely to maximize the final answer quality.

\begin{figure*}[htbp]
    \centering
    \includegraphics[width=\linewidth]{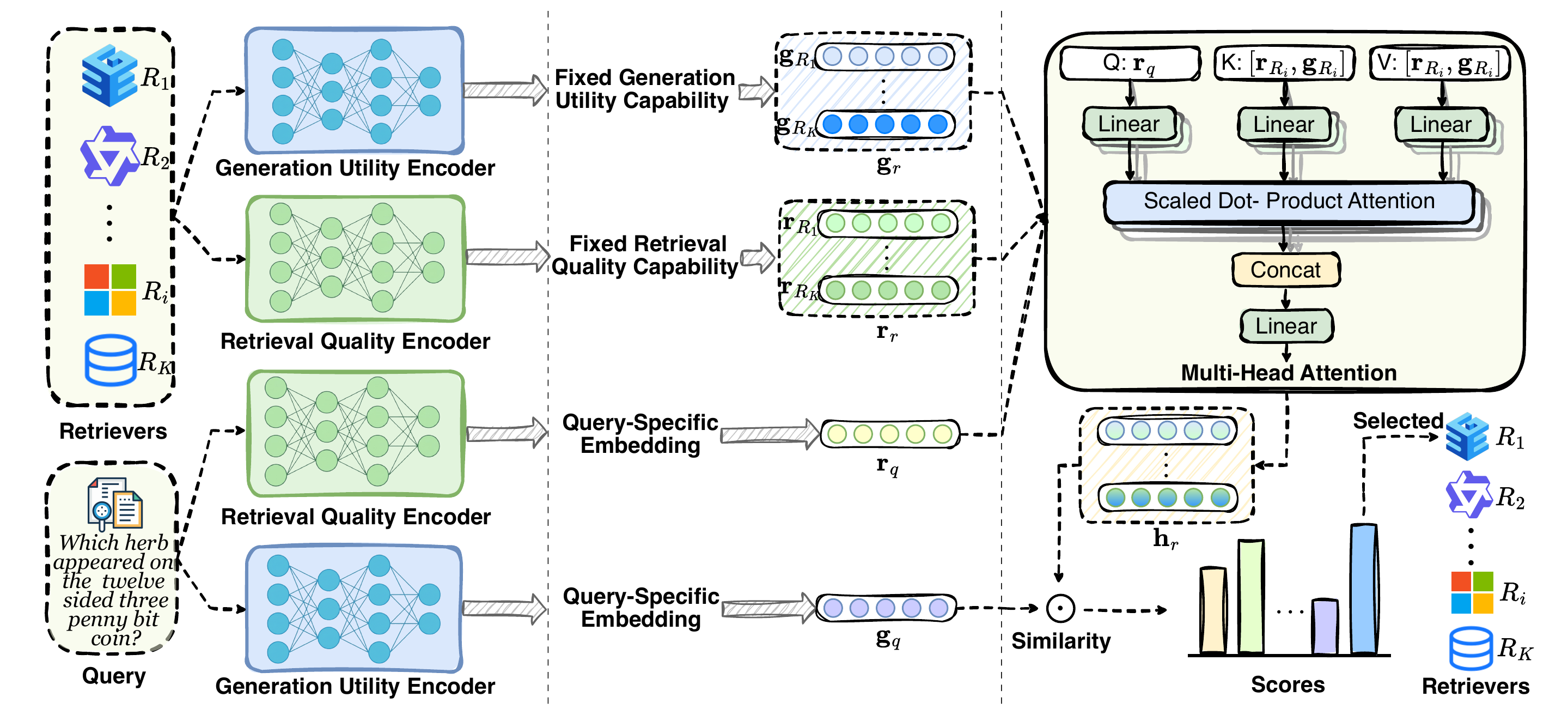}
    \caption[The inference of \ours{}]{\ours{} models retriever capability from retrieval quality and generation utility, and applies query-adaptive multi-head attention to select the most suitable retriever for each query.}
    \label{inference}
\end{figure*}

\section{Methodology}
We introduce \ours{}, a query-adaptive routing framework for RAG scenarios. As introduced in Section~\ref{sec:intro}, effective routing requires modeling not just retrieval \textbf{relevance} but also the \textbf{utility} of retrieved documents for downstream generation. To capture this duality, \ours{} disentangles retriever capability into two distinct dimensions: \textit{Retrieval Quality} and \textit{generation utility}.

The overall framework is illustrated in Figure~\ref{inference}. We first describe the architecture of \ours{}, which consists of dual-encoder capability modeling, query-conditioned capability fusion, and similarity-based routing (Section~\ref{sec:arch}). Subsequently, we detail our two-stage optimization strategy, which employs contrastive learning to align these capability representations with supervision signals derived from both document assessments and generation outcomes (Section~\ref{sec:opt}).

\subsection{Architecture}\label{sec:arch}
The architecture of \ours{} is designed to transform raw inputs into actionable routing decisions through a structured pipeline. Specifically, the model first maps the query and candidates into a disentangled capability space, then fuses these representations based on the query's specific needs, and finally computes a similarity score for routing. This process comprises three key components: capability representation, capability fusion, and the final routing decision.


\paragraph{Capability Representation}
We construct a representational space where both queries and retrievers are characterized by two complementary factors:
(1) \textbf{Retrieval Quality~($\mathbf{r}$)}, which captures intrinsic retrieval properties such as relevance, coverage, and redundancy; and
(2) \textbf{generation utility~($\mathbf{g}$)}, which reflects the potential to support correct downstream answers.

To implement this, \ours{} employs two shared encoders: a {retrieval quality encoder} $\phi_r$ and a {generation utility encoder} $\phi_g$.
Given a query $q$ and a retriever $R_i$, we obtain their respective capability embeddings as:
\begin{equation}
\begin{aligned}
    \mathbf{r}_q &= \phi_r(q), & \mathbf{g}_q &= \phi_g(q), \\
    \mathbf{r}_{R_i} &= \phi_r(R_i), & \mathbf{g}_{R_i} &= \phi_g(R_i).
\end{aligned}
\end{equation}
Here, the pair $\{\mathbf{r}_{R_i}, \mathbf{g}_{R_i}\}$ constitutes the comprehensive capability representation for the retriever $R_i$. Crucially, encoders of the same type share parameters across queries and retrievers to ensure a consistent metric space.




\paragraph{Capability Fusion}
Since different queries prioritize different retriever capabilities, a static combination of $\mathbf{r}_{R_i}$ and $\mathbf{g}_{R_i}$ is insufficient. \ours{} employs a multi-head attention mechanism to dynamically fuse these capabilities conditioned on the query's retrieval intent $\mathbf{r}_q$.
Formally, the fused retriever representation $\mathbf{h}_{R_i}$ is computed as:
\begin{equation}
\mathbf{h}_{R_i} =
\operatorname{MHA}\!\left(
    \mathbf{r}_q,\;
    [\mathbf{r}_{R_i}, \mathbf{g}_{R_i}],\;
    [\mathbf{r}_{R_i}, \mathbf{g}_{R_i}]
\right),
\end{equation}
where $\operatorname{MHA}(\cdot)$ denotes the multi-head attention operation and $[\cdot,\cdot]$ indicates vector concatenation. This mechanism allows the router to weigh retrieval quality and generation utility differently depending on the specific retrieval needs of the query $q$.


\paragraph{Similarity-Based Retriever Routing}
The final routing result is based on the alignment between the query's generation requirement $\mathbf{g}_q$ and the fused retriever capability $\mathbf{h}_{R_i}$. We select the optimal retriever $R_{\pi(q)}$ by maximizing the cosine similarity:
\begin{equation}
\pi(q) = \arg\max_{R_i \in \mathcal{R}}
\operatorname{sim}(\mathbf{g}_q, \mathbf{h}_{R_i}),
\end{equation}
where $\operatorname{sim}(\cdot,\cdot)$ denotes cosine similarity. This formulation ensures that the selected retriever is the one whose fused capability best matches the generation utility required by the query. This design choice reflects our emphasis on the final generation quality, as the retriever is selected based on how well it supports the downstream generation rather than retrieval performance alone.

\subsection{Optimization}\label{sec:opt}
The effectiveness of a retriever in RAG is dynamic, exhibiting query-specific shifts in both retrieval quality and generation utility. To effectively capture these shifts, we employ a two-stage optimization strategy. In the first stage, we train the capability encoders ($\phi_r, \phi_g$) using contrastive learning with disentangled supervision signals. In the second stage, we freeze these encoders and optimize the fusion mechanism to learn the optimal combination of capabilities.



\subsubsection{Optimizing Capability Encoders}
We design two complementary contrastive objectives to train $\phi_r$ and $\phi_g$ separately. This isolation ensures that each encoder captures its specific aspect of capability without interference from the other.

\paragraph{Retrieval Quality Encoder ($\phi_r$)}
For retrieval quality, supervision should be independent of the generator to avoid confounding errors. We employ a powerful LLM (\textit{e.g.}, Qwen3-Next-80B-A3B-Instruct~\citep{DBLP:journals/corr/abs-2505-09388}) to assess retrieved documents along dimensions such as relevance, coverage, and redundancy; the details are provided in Appendix~\ref{sec:retrieval_quality_prompt}. Let $s(q, R_i)$ be the aggregated retrieval quality score.
We construct positive set $\mathcal{V}^+_r(q)$ and negative set $\mathcal{V}^-_r(q)$ based on the top-$k$ ranking of $s(q, R_i)$:
\begin{equation}
\left\{
\begin{aligned}
\mathcal{V}^+(q) &= \{\, \mathbf{r}_{R_i} \mid R_i \in \operatorname{Top}\text{-}k_{R}(s(q, R)) \,\}, \\
\mathcal{V}^-(q) &= \{\, \mathbf{r}_{R_i} \mid R_i \in \mathcal{R} \setminus
\operatorname{Top}\text{-}k_{R}(s(q, R)) \,\}.
\end{aligned}
\right.
\notag
\end{equation}
The corresponding contrastive loss is formulated as:
\begin{align}
\mathcal{L}_{\text{qual}}(q) &= \sum_{\mathbf{r}^+ \in \mathcal{U}^+(q)}
\Bigl(
- \tfrac{1}{\tau}\operatorname{sim}(\mathbf{r}_q, \mathbf{r}^+)
+ \log Z_{\text{r}}(q)
\Bigr),\notag \\
Z_{\text{r}}(q) &= \sum_{\mathbf{r}\in \mathcal{V}^+(q)\cup \mathcal{V}^-(q)}
\exp\bigl(\operatorname{sim}(\mathbf{r}_q,\mathbf{r})/\tau\bigr),\notag
\end{align}
where $\operatorname{sim}(\cdot,\cdot)$ denotes cosine similarity, $\tau$ is a temperature hyperparameter, and we set $k=2$ in all experiments.

\paragraph{Generation Utility Encoder ($\phi_g$)} 

For generation utility modeling, we derive supervision signals from downstream generation outcomes. However, relying solely on the correctness of a single generated answer can be noisy and insufficient to characterize overall generation utility. To obtain a more robust supervision signal, we consider both query-level and retriever-level generation performance. Specifically, for a given query $q$ and retriever $R_i$, we compute a generation utility score $u(q, R_i)$ combining: (i) query-level exact match (EM), (ii) query-level F1 score, and (iii) the retriever's global average correctness $\sigma(R_i)$:
\begin{equation}
u(q, R_i) = \text{EM}(q, R_i) + \beta \cdot \text{F1}(q, R_i) + \gamma \cdot {\sigma}(R_i).
\end{equation}
where $\beta$, and $\gamma$ are weighting coefficients.

Similar to retrieval quality, we form positive set $\mathcal{V}^+_g(q)$ and negative set $\mathcal{V}^-_g(q)$ based on the top-$k$ ranking of $u(q, R_i)$:
\begin{equation}
\left\{
\begin{aligned}
\mathcal{V}^+(q) &= \{\, \mathbf{g}_{R_i} \mid R_i \in \operatorname{Top}\text{-}k_{R}(u(q, R)) \,\} \\
\mathcal{V}^-(q) &= \{\, \mathbf{g}_{R_i} \mid R_i \in \mathcal{R} \setminus
\operatorname{Top}\text{-}k_{R}(u(q, R)) \,\}.
\end{aligned}
\right.\notag
\end{equation}
The corresponding contrastive loss is defined as:
\begin{align}
\mathcal{L}_{\text{util}}(q)&= \sum_{\mathbf{g}^+ \in \mathcal{V}^+(q)}
\Bigl(
- \tfrac{1}{\tau}\operatorname{sim}(\mathbf{g}_q, \mathbf{g}^+)
+ \log Z_{\text{g}}(q)
\Bigr),\notag\\
Z_{\text{g}}(q)&=\sum_{\mathbf{g}\in \mathcal{V}^+(q)\cup \mathcal{V}^-(q)}
\exp\!\bigl(\operatorname{sim}(\mathbf{g}_q,\mathbf{g})/\tau\bigr).\notag
\end{align}

\begin{figure}[t]
    \centering
    \includegraphics[width=\linewidth]{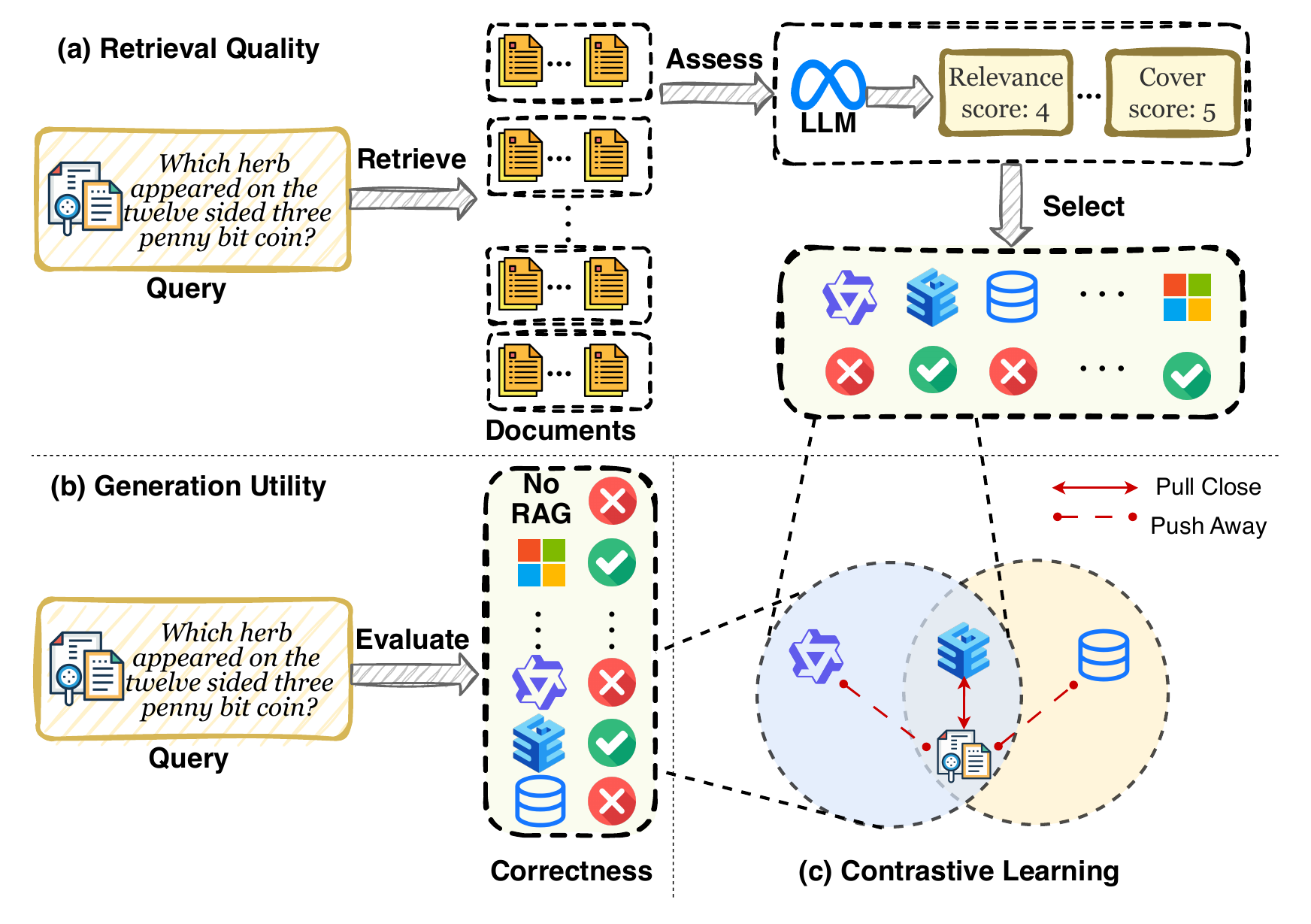}
    \caption{(a) Retrieval Quality supervision provides contrastive signals based on document-level quality assessments, decoupled from generation outcomes. (b) Generation Utility supervision derives labels from downstream answer quality, including both retrieval and no-retrieval cases. (c) Contrastive learning aims to align positive samples more closely with the query representation while distancing negative ones from it.}
    \label{fig:sample}
\end{figure}

\subsubsection{Optimizing Capability Fusion}
Once the capability encoders are trained, we freeze their parameters and focus on optimizing the multi-head attention module. The goal of this stage is to learn how to weigh retrieval quality and generation utility dynamically to maximize the likelihood of selecting a useful retriever.

We define a matching probability $p(q, R_i)$ based on the similarity between the fused representation $\mathbf{h}_{R_i}$ and the query's utility requirement $\mathbf{g}_q$:
\begin{equation}
p(q, R_i) = \operatorname{sigmoid}\big( \operatorname{sim}(\mathbf{h}_{R_i}, \mathbf{g}_q) \big).
\end{equation}
The optimization target is to predict whether a retriever is ``useful'' (\textit{i.e.}, belongs to the top-$k$ positive set defined in the Generation Utility Encoder optimization). Let $y_{q_i} \in \{0, 1\}$ be the binary label indicating if $R_i$ is a positive sample for $q$. The classification loss is:
\begin{equation}
\mathcal{L}_{\text{cls}}
= \mathbb{E}_{q} \sum_{R_i} \operatorname{BCE}\big(p(q, R_i), y_{q_i}\big).
\end{equation}
where $\operatorname{BCE}(\cdot,\cdot)$ denotes the binary cross-entropy loss.

Additionally, we apply a regularization term to prevent the fused representation from deviating excessively from the generation utility embedding and ensure stability:
\begin{equation}
\mathcal{R}(\mathbf{h}_{R_i}, \mathbf{g}_{R_i}) = \left\| \mathbf{h}_{R_i} - \mathbf{g}_{R_i} \right\|_2^2 .
\end{equation}

The final objective for this stage is:
\begin{equation}
\mathcal{L}
= \mathcal{L}_{\text{cls}}
+ \lambda_{\text{reg}} \, \mathcal{R}(\mathbf{h}_{R_i}, \mathbf{g}_{R_i}).
\end{equation}

\section{Experiments}
\subsection{Experimental Setup}
\paragraph{Benchmarks}
To comprehensively evaluate \ours{} across varying levels of retrieval complexity, we conduct experiments on three representative knowledge-intensive benchmarks:
(1)~\textbf{TriviaQA}~\citep{DBLP:conf/acl/JoshiCWZ17}, a large-scale open-domain reading comprehension dataset featuring factoid questions collected from trivia enthusiasts and paired with evidence from web documents;
(2)~\textbf{Natural Questions (NQ)}~\citep{DBLP:journals/tacl/KwiatkowskiPRCP19}, a benchmark consisting of real-world search queries that typically require identifying short answers within Wikipedia articles; and
(3)~\textbf{HotpotQA}~\citep{DBLP:conf/emnlp/Yang0ZBCSM18}, a challenging dataset that requires multi-hop reasoning across multiple supporting documents to derive the final answer.
Following previous studies~\cite{DBLP:conf/www/Jin0DDYZZYW25}, we apply Exact Match (EM) and F1 score as the evaluation metrics.

\paragraph{Candidate Retrievers} 
To simulate a realistic and heterogeneous retrieval environment, we construct a candidate pool of eight widely used retrievers. This set includes the classical sparse retriever BM25 and seven dense retrievers, spanning a broad spectrum of model sizes from 0.1B to 4B parameters. Comprehensive statistics on model scale and inference latency are reported in Table~\ref{tab:retriever_params_latency}, while detailed implementation specifications are provided in Appendix~\ref{sec:details_of_retrievers}.

\begin{table}[t]
\centering
\small
\caption{Model size and latency of different retrievers. BM25 does not contain trainable parameters.}
\label{tab:retriever_params_latency}
\setlength{\tabcolsep}{2pt}
\begin{tabular}{p{0.5\linewidth}cc}
\toprule
\textbf{Retriever} & \textbf{Param. (B)} & \textbf{Latency (ms)} \\
\midrule
E5-base-v2          & 0.11 & 6.40 \\
E5-large            & 0.34 & 10.11 \\
BGE-large           & 0.34 & 11.09 \\
BGE-m3             & 0.57 & 11.42 \\
DIVER-Retriever-0.6B    & 0.60 & 23.39 \\
Qwen3-embedding-0.6B    & 0.59 & 25.11 \\
Qwen3-embedding-4B      & 4.02 & 32.19 \\
\midrule
BM25 (warm)         & --   & 43.08 \\
BM25 (cold)         & --   & 91.49 \\
\bottomrule
\end{tabular}
\end{table}

\begin{table*}[t]
\centering
\small
\caption{Performance comparison of \ours{} against retriever-based, rule-based, and adaptive RAG baselines across various knowledge-intensive tasks and RAG settings. EM and F1 scores on the test set are reported. ``Pipeline Type'' denotes the generation control paradigm. The best results are highlighted in \textbf{bold}, and the second best results are \underline{underlined}.}
\label{tab:retriever_tqa_nq_hotpot}
\setlength{\tabcolsep}{2mm}
\begin{tabular}{p{0.25\textwidth} c c c c c c c c c}
\toprule
\multirow{2}{*}{\textbf{Methods}} 
& \multirow{2}{*}{\textbf{Pipeline Type}}
& \multicolumn{2}{c}{\textbf{TriviaQA}} 
& \multicolumn{2}{c}{\textbf{NQ}} 
& \multicolumn{2}{c}{\textbf{HotpotQA}} 
& \multicolumn{2}{c}{\textbf{Average}}\\
\cmidrule(lr){3-4}\cmidrule(lr){5-6}\cmidrule(lr){7-8}\cmidrule(lr){9-10}
 &  & \textbf{EM} & \textbf{F1} 
 & \textbf{EM} & \textbf{F1} 
 & \textbf{EM} & \textbf{F1} 
 & \textbf{EM}  & \textbf{F1}\\
\midrule
Naive            & -- & 55.34 & 61.81 & 22.35 & 31.41 & 19.16 & 26.96 & 32.28 & 40.06\\
E5-base-v2       & Sequential & 58.83 & 68.12 & 36.12 & 47.35 & 25.31 & 35.75 & 40.09 & 50.41\\
E5-large         & Sequential & 60.16 & 69.56 & 36.89 & 48.00 & 26.87 & 37.87 & 41.31 & 51.81\\
BGE-large        & Sequential & 57.02 & 66.30 & 34.51 & 45.27 & 27.54 & 38.60 & 39.69 & 50.06\\
BGE-m3           & Sequential & 55.15 & 64.21 & 33.52 & 43.49 & 24.73 & 35.01 & 37.80 & 47.57\\
Qwen3-embedding-0.6b & Sequential & 55.89 & 65.33 & 32.35 & 42.71 & 24.42 & 34.63 & 37.55 & 47.56\\
Qwen3-embedding-4b   & Sequential & 59.75 & 69.38 & 34.49 & 45.80 & 26.64 & 37.62 & 40.29 & 50.93\\
DIVER-Retriever-0.6b & Sequential & 55.56 & 64.76 & 30.66 & 40.36 & 22.75 & 32.78 & 36.32 & 45.97\\
BM25             & Sequential & 48.70 & 59.04 & 12.93 & 18.15 & 17.87 & 26.15 & 26.50 & 34.45\\
\midrule
Random            & Conditional & 56.49 & 65.60 & 30.53 & 40.51 & 22.13 & 32.09 & 36.38 & 46.07\\
Oracle Single Best& Conditional & 60.16 & 69.56 & \underline{36.89} & \underline{48.00} & 27.54 & 38.60 & 41.53 & 52.05\\
\midrule
FLARE            & Loop & 56.73 & 66.01 & 22.51 & 32.42 & 20.77 & 28.03 & 33.34 & 42.15\\
IRCoT            & Loop & \underline{60.71} & \underline{69.92} & 35.81 & 47.95 & \underline{29.05} & 40.52 & \underline{41.86} & \underline{52.80}\\
Adaptive-RAG     & Conditional & 57.41 & 66.81 & 35.12 & 45.55 & 28.87 & \underline{40.54} & 40.47 & 50.97\\
AAR-contriever-kilt & Sequential & 58.82 & 67.23 & 30.12 & 40.44 & 23.04 & 33.37 & 37.33 & 47.01\\
LTRR & Route & 58.79 & 68.00 & 35.76 & 46.90 & 19.86 & 28.64 & 39.14 & 48.85 \\
RouterRetriever  & Route & 57.42 & 66.73 & 32.44 & 43.31 & 24.37 & 34.98 & 38.08 & 48.34 \\
\midrule
\textbf{\ours{} (ours)} 
& Route
& \textbf{62.24} & \textbf{71.52} 
& \textbf{37.81} & \textbf{49.32} 
& \textbf{29.13} & \textbf{40.73} & \textbf{43.06} & \textbf{53.86} \\
\bottomrule
\end{tabular}
\end{table*}

\paragraph{Settings} 

We employ \texttt{LLaMA3-8B-Instruct} as the backbone generator~\citep{DBLP:journals/corr/abs-2407-21783}, configured with a maximum context window of 2,048 tokens. For the knowledge source, we use the 2018 English Wikipedia dump~\citep{DBLP:conf/emnlp/KarpukhinOMLWEC20}, retrieving the top-5 documents per query. For baselines utilizing a single fixed retriever, we select \texttt{E5-large} by default, as it demonstrated the best average performance in our preliminary analysis. To ensure fair comparison, standardized prompts are employed for all RAG and non-RAG settings, respectively, with full details provided in Appendix~\ref{sec:global_setting}. 

To evaluate the effectiveness of different retrieval paradigms, we categorize the compared methods into four distinct types:
(1) {\textit{Sequential}}: Standard RAG approaches that follow a fixed ``retrieve-then-generate'' pipeline using a single retriever;
(2) {\textit{Conditional}}: Adaptive methods that explicitly decide whether to perform retrieval based on the input query;
(3) {\textit{Loop}}: Iterative methods that interleave retrieval and generation steps, dynamically determining {when} and {what} to retrieve during inference; and
(4) {\textit{Route}}: Routing-based methods, including our approach, which dynamically select the most suitable retriever from a candidate pool for each query.

\paragraph{Supervision Cost and Reproducibility.} Since \ours{} relies on both retrieval-quality and generation-utility supervision, we report the cost in a throughput-based form. Specifically, generation-utility supervision is constructed from answer generation over $(q, R_i)$ pairs, while retrieval-quality supervision is constructed using an external LLM judge over retrieved documents. We report measured throughput and approximate cost per 10k items in Appendix~\ref{app:supervision_cost} for easier scaling to different budgets. 

\paragraph{Baselines} We compare \ours{} against a comprehensive set of baselines organized into three categories. 
(1) \textbf{ Individual Retrievers and Naive Generation:}
We first report the performance of the underlying generator without retrieval (\textbf{Naive}) and combined with each of the eight candidate retrievers individually (e.g., BM25, E5, etc., as detailed in the previous section). These results establish the fundamental performance spectrum of standard RAG systems.
(2) \textbf{Heuristic Routing Strategies:}
To assess the difficulty of the routing task, we provide reference points using rule-based strategies:
(i) \textbf{Random}, which selects a retriever uniformly at random for each query; and
(ii) \textbf{Oracle Single Best}, which ideally selects the best-performing retriever on each dataset.
(3) \noindent\textbf{RAG Baselines:}
Finally, we compare against advanced methods that optimize the RAG workflow:
(i) \textbf{FLARE}~\citep{DBLP:conf/emnlp/JiangXGSLDYCN23} and (ii) \textbf{IRCoT}~\citep{DBLP:conf/acl/TrivediBKS23}, loop-based methods that interleave retrieval and generation, utilizing active confidence checks or chain-of-thought reasoning to guide retrieval;
(iii) \textbf{Adaptive RAG}~\citep{DBLP:conf/naacl/JeongBCHP24}, a conditional method that dynamically routes queries to no-retrieval, single-step, or multi-step pipelines based on complexity;
(iv) \textbf{AAR}~\citep{DBLP:conf/acl/YuXY023}, which optimizes the retriever to align with LLM preferences, serving as a strong single-retriever baseline;
(v) \textbf{RouterRetriever}~\citep{DBLP:conf/aaai/LeeSCSL25}, a routing method designed for IR settings that selects among domain-specific experts; and
(vi) \textbf{LTRR}~\citep{DBLP:journals/corr/abs-2506-13743}, which formulates retriever routing as a learning-to-rank problem to dynamically rank candidate retrievers.

\newcommand{\yes}{\textcolor{teal}{\ensuremath{\checkmark}}}
\newcommand{\no}{\textcolor{red!85!black}{\ensuremath{\times}}}

\begin{table*}[t]
\centering
\small
\caption{Ablation study of \ours{} components.
$\Delta$EM and $\Delta$F1 denote the average performance change across TQA, NQ, and HQA relative to the full model. RQ Encoder and GU Encoder denote the Retrieval Quality Encoder and Generation Utility Encoder, respectively.}
\label{tab:ablation_router_r3_component}
\setlength{\tabcolsep}{1.5mm}
\begin{tabular}{lccc c c c c c c c c}
\toprule
\multirow{2}{*}{\textbf{}}
& \multirow{2}{*}{\textbf{MHA}}
& \multirow{2}{*}{\textbf{RQ Encoder}}
& \multirow{2}{*}{\textbf{GU Encoder}}
& \multicolumn{2}{c}{\textbf{TQA}}
& \multicolumn{2}{c}{\textbf{NQ}}
& \multicolumn{2}{c}{\textbf{HQA}}
& \multirow{2}{*}{$\boldsymbol{\Delta}$\textbf{EM}}
& \multirow{2}{*}{$\boldsymbol{\Delta}$\textbf{F1}} \\
\cmidrule(lr){5-6}\cmidrule(lr){7-8}\cmidrule(lr){9-10}
 & &  &  & \textbf{EM} & \textbf{F1}
 & \textbf{EM} & \textbf{F1}
 & \textbf{EM} & \textbf{F1}
 &  &  \\
\midrule
\textbf{\ours{}}&
\yes & \yes & \yes
& 62.24 & 71.52
& 37.41 & 48.72
& 29.05 & 41.33
& 0.00 & 0.00 \\
\quad w/o MHA&
\no & \yes & \yes
& 61.57 & 70.43
& 36.87 & 47.94
& 27.58 & 38.76
& -0.89 & -1.88 \\
\quad w/o RQ Encoder&
\no & \yes & \no
& 60.95 & 69.03
& 36.91 & 48.13
& 28.03 & 38.82
& -0.93 & -2.26 \\
\quad w/o GU Encoder&
\no & \no & \yes
& 61.80 & 70.77
& 36.83 & 47.56
& 27.05 & 38.00
& -1.01 & -2.15 \\
\bottomrule
\end{tabular}
\end{table*}

\begin{figure*}[t]
    \centering
    \includegraphics[width=\linewidth]{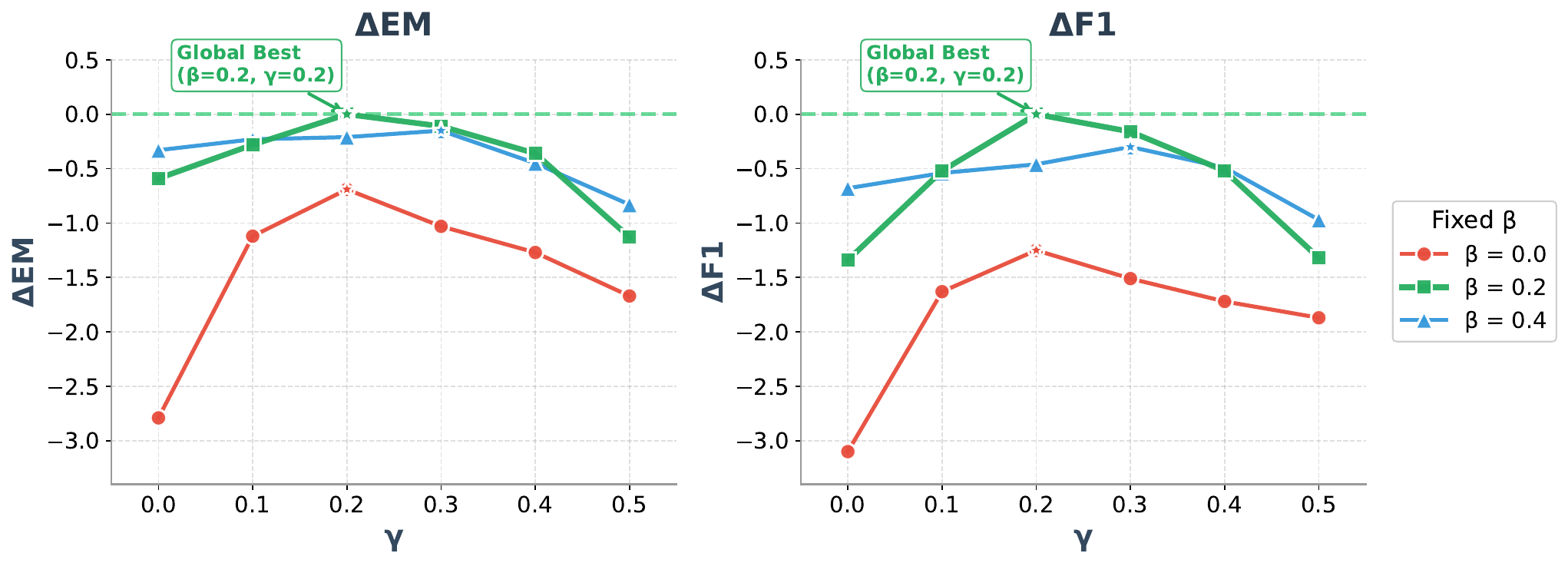}
    \caption{Sensitivity analysis of the $\beta$ and $\gamma$ weighting factors in training the Generation Utility Encoder.}
    \label{fig:ablation}
\end{figure*}






\subsection{Overall Results}
\paragraph{Comparison with Single Non-Routed Retrievers.} As shown in Table~\ref{tab:retriever_tqa_nq_hotpot}, \ours{} attains the strongest test-set performance across diverse knowledge-intensive benchmarks and RAG configurations, reaching 43.06 average EM and 53.86 average F1. Compared with the top performed standalone retriever, E5-large, which obtains 41.31 EM and 51.81 F1, \ours{} yields clear gains, underscoring the benefit of routing. In addition, \ours{} exceeds the Oracle Single Best baseline, which chooses the most effective individual retriever for each dataset and achieves 41.53 EM and 52.05 F1. This comparison indicates that dynamically routing among multiple retrievers can better exploit their complementary capabilities and achieve results unattainable by any single retriever alone.

\paragraph{Comparison with Single and Static routing baselines.} \ours{} significantly outperforms single and static routing baselines, for example, RouterRetriever (38.08 and 48.34) and LTRR (39.14 and 48.85). These findings highlight the limitations of approaches that do not explicitly capture retriever capability, including both retrieval quality and generation utility, and that are unable to adapt to the evolving interaction between the retriever and the query. Moreover, \ours{} also surpasses rule-based strategies such as Random (36.38 and 46.07) and Oracle Single Best (41.53 and 52.05). Taken together, these results substantiate our central claim that accurately routing retrievers requires explicitly modeling both retrieval quality and generation utility.

\paragraph{Comparison with Retrieval Necessity Judgement Approaches.} \ours{} can also handle the task of retrieval necessity judgement. To reduce unnecessary retrievals, this problem has been extensively studied in the RAG literature. However, existing approaches typically focus on deciding whether to invoke a \emph{single} retriever, treating retrieval as a binary decision. Such formulations can be viewed as a special case of retriever routing, which is naturally subsumed by \ours{}. Empirically, \ours{} outperforms these studies almost consistently, including FLARE (33.34 and 42.15), IRCOT (41.86 and 52.80), and Adaptive-RAG (40.47 and 50.97). These results suggest that retriever routing offers a principled generalization of retrieval necessity judgment in RAG settings.

\begin{table*}[t]
\centering
\small
\caption{Analysis of $R_0$ (no-retrieval) selection.
EM@$R_0$ denotes the exact match score on the subset where the router selects $R_0$.}
\label{tab:r0_analysis}
\setlength{\tabcolsep}{2.5mm}
\begin{tabular}{lcccc}
\toprule
\textbf{Dataset} & \textbf{$R_0$ Selection Rate (\%)} & \textbf{EM@$R_0$ (\%)} & \textbf{No-Retrieval EM (\%)} & \textbf{\ours{} EM (\%)} \\
\midrule
TriviaQA & 37.82 & 76.33 & 55.34 & 62.24 \\
NQ       & 15.23 & 43.51 & 22.35 & 37.81 \\
HotpotQA & 9.52  & 31.78 & 19.16 & 29.13 \\
\bottomrule
\end{tabular}
\end{table*}

\begin{table*}[t]
\centering
\small
\caption{Cross-generator evaluation with Qwen3-8B at test time.
The router is trained with LLaMA3-8B-Instruct and directly transferred without retraining.}
\label{tab:cross_generator}
\setlength{\tabcolsep}{3mm}
\begin{tabular}{lcccccccc}
\toprule
\multirow{2}{*}{\textbf{Methods}} 
& \multicolumn{2}{c}{\textbf{TriviaQA}} 
& \multicolumn{2}{c}{\textbf{NQ}} 
& \multicolumn{2}{c}{\textbf{HotpotQA}} 
& \multicolumn{2}{c}{\textbf{Average}} \\
\cmidrule(lr){2-3}\cmidrule(lr){4-5}\cmidrule(lr){6-7}\cmidrule(lr){8-9}
& \textbf{EM} & \textbf{F1} 
& \textbf{EM} & \textbf{F1} 
& \textbf{EM} & \textbf{F1} 
& \textbf{EM} & \textbf{F1} \\
\midrule
Naive                & 52.97 & 62.98 & 22.44 & 35.87 & 19.15 & 29.73 & 31.52 & 42.86 \\
Random               & 53.57 & 64.98 & 26.41 & 37.22 & 29.09 & 40.06 & 36.36 & 47.42 \\
Oracle-Single-Best   & 59.45 & 69.70 & 31.86 & 44.98 & 32.49 & 44.41 & 41.27 & 53.03 \\
\ours{}              & 60.11 & 70.48 & 30.72 & 43.65 & 31.62 & 44.34 & 40.82 & 52.82 \\
\bottomrule
\end{tabular}
\end{table*}

\subsection{Ablation Study and Sensitivity Analysis} \label{sec:ablation}
\paragraph{Effects of \ours{} Architecture.} We conduct an ablation study to evaluate the contributions of the Retrieval Quality (RQ) Encoder, Generation Utility (GU) Encoder, and the Multi-Head Attention (MHA) mechanism. As illustrated in Table~\ref{tab:ablation_router_r3_component}, removing MHA and simply averaging retriever representations leads to consistent performance degradation across all datasets, highlighting the importance of query-adaptive capability fusion. Ablating either the RQ Encoder or the GU Encoder also results in noticeable performance drops, indicating that retrieval quality and generation utility provide complementary signals for effective routing. The full \ours{} model, which jointly models both factors and integrates them via MHA, achieves the best performance across all benchmarks.

\paragraph{Effects of Different Parameters in Sampling.} We justify the choice of weighting factors $\beta$ and $\gamma$ through a sensitivity analysis shown in Figure~\ref{fig:ablation}.  The results show that combining both signals consistently outperforms using either one alone, while removing both leads to substantial performance degradation. Performance peaks at moderate values of $\beta$ and $\gamma$ (around 0.2). This trend indicates that retriever-level and answer-level supervision provide complementary but non-redundant signals, and balanced weighting is essential for stable training and effective retriever routing. Based on these observations, we set $\beta=\gamma=0.2$ in all experiments.

\paragraph{Analysis of $R_0$ Selection.} We further analyze the router's use of the no-retrieval option $R_0$. As shown in Table~\ref{tab:r0_analysis}, $R_0$ is selected in 37.82\%, 15.23\%, and 9.52\% of cases on TriviaQA, NQ, and HotpotQA, respectively, indicating that a non-trivial portion of queries can be handled without retrieval. In addition, EM@$R_0$ is consistently much higher than the overall no-retrieval baseline on all datasets, suggesting that the router can reliably identify queries for which retrieval is unnecessary. These results confirm that $R_0$ functions as an effective selective abstention option rather than a trivial fallback.

\paragraph{Analysis of Cross-Generator Generalization.} We further evaluate whether the learned routing policy transfers across generators. In our main setting, the router is trained with LLaMA3-8B-Instruct as the backbone generator. To test robustness under a backbone swap, we directly replace the generator with Qwen3-8B at test time, while keeping the router fixed and performing no retraining. As shown in Table~\ref{tab:cross_generator}, \ours{} remains robust and competitive under this setting, achieving the best performance on TriviaQA and strong average results across the three datasets. This result suggests that, although generation utility may vary across generators, the learned routing policy transfers reasonably well across strong instruction-tuned LLMs. We attribute this robustness partly to our disentangled design, where the retrieval-quality channel provides a more stable, generator-agnostic signal.

\section{Conclusion}

In this paper, we explore the problem of retriever routing in RAG and propose \ours{}, a query-adaptive routing framework that goes beyond the conventional assumption of static retriever capability. Rather than treating retrievers as interchangeable modules or assessing them solely by retrieval relevance, \ours{} explicitly models two complementary aspects of retriever capability: retrieval quality and generation utility. By disentangling these factors and fusing them in a query-conditioned manner, \ours{} captures query-specific preferences over candidate retrievers. Experiments on multiple knowledge-intensive benchmarks show that \ours{} consistently outperforms strong single-retriever baselines, static routing methods, and retrieval necessity judgment approaches. Additional analyses further confirm that both retrieval quality and generation utility are crucial for effective routing. Overall, our results highlight the value of capability-aware retriever selection for building more adaptive and effective RAG systems.

\section{Limitations}
Despite achieving strong performance across multiple tasks and effectively fulfilling the intended routing objective, there remain several aspects that could be further explored to enhance the generality and applicability of the framework. First, motivated by lightweight design considerations, the current approach adopts relatively simple representations for capability modeling; future work may investigate whether more expressive architectures can yield additional performance gains while maintaining a favorable efficiency--effectiveness balance. Second, although the router is able to select appropriate retrievers for individual queries, it does not yet consider more complex retrieval scenarios that involve combining multiple retrievers, which represents another promising direction for future exploration. Finally, the experimental evaluation is currently limited to question answering tasks, and extending the framework to a broader range of task settings would help better assess its generality.


\bibliography{custom}

\appendix

\section{Experimental Setup Details} \label{sec:appendix}
\subsection{Details of Retrievers} \label{sec:details_of_retrievers}
\begin{figure*}[t]
    \centering
    \includegraphics[width=\linewidth]{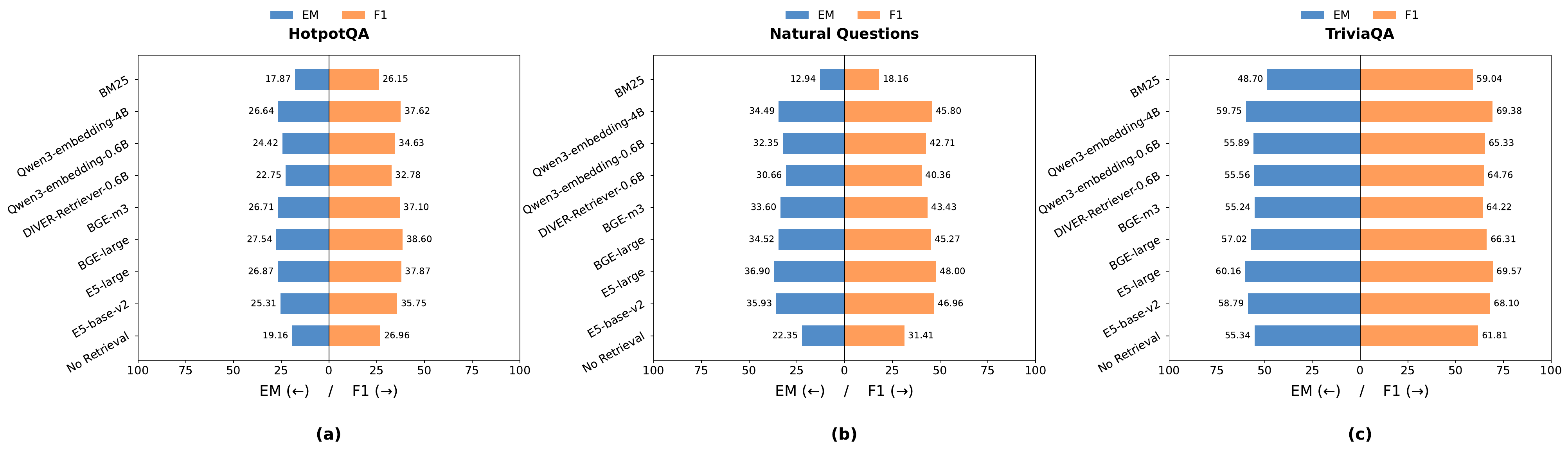}
    \caption{Mirrored horizontal bars show EM (left) and F1 (right) on HotpotQA, Natural Questions, and TriviaQA. Results are reported for multiple retrievers and a no-retrieval baseline, evaluated on the set of questions shared across all retrieval settings.}
    \label{fig:three_datasets_em_f1}
\end{figure*}
All candidate retrievers are evaluated under a unified and reproducible experimental setup. Dense retrievers are locally deployed using the Faiss backend with GPU acceleration, while the sparse retriever BM25 is implemented via Pyserini. For dense models, inference latency is measured as the average encoding and retrieval time per query, excluding data loading and preprocessing overhead.

The candidate pool consists of one sparse retriever (BM25) and seven dense retrievers with parameter sizes ranging from 0.11B to 4.02B, covering a broad spectrum of efficiency--accuracy trade-offs commonly encountered in practical RAG systems. All dense retrievers are used with their publicly released checkpoints, without additional fine-tuning, to ensure that observed performance differences stem from inherent model capability rather than task-specific adaptation. For BM25, we report both warm and cold latency to reflect different deployment conditions. Warm latency measures retrieval time with the index already loaded in memory, while cold latency additionally includes index loading overhead. In contrast, latency measurements for dense retrievers only account for model inference and nearest-neighbor search, as their indices remain resident in GPU memory during evaluation.

Comprehensive statistics on model size and inference latency are reported in Table~\ref{tab:retriever_params_latency}. In addition, we provide per-dataset evaluation results for each retriever in the Figure~\ref{fig:three_datasets_em_f1}, including EM and F1 scores on TriviaQA, NQ, and HotpotQA, to enable
a more fine-grained comparison of retriever behavior across datasets.

\subsection{Details of Global Setting} \label{sec:global_setting}
For the eight retrievers evaluated in this work, as well as other methods, unless
explicitly stated otherwise, we adopt a standard retrieve--then--generate
paradigm with a unified prompt template.
Specifically, the retrieved documents are directly concatenated with the input
query and fed into the generator, without any additional routing, filtering, or
iterative control mechanisms.
The system prompt instructs the model to generate answers strictly grounded in
the provided documents:
\begin{quote}
\small
\texttt{Answer the question based on the given document. Only give me the answer and do not output any other words. The following are given documents:}
\end{quote}

In the non-RAG setting, the prompt is defined as:
\begin{quote}
\small
\texttt{Answer the question based on your own knowledge. Only give me the answer and do not output any other words.}
\end{quote}

The retrieved documents are appended in a structured format (\texttt{Doc $i$ (Title:\{title\}) \{content\}}), where the top-$5$ documents are used for each query, followed by the user prompt \texttt{Question:\{question\}}. The system prompt, retrieved documents, and user prompt are combined using the \texttt{tokenizer.apply\_chat\_template} function to construct the final input sequence for the generator. This unified prompt design is consistently applied across all Naive RAG baselines to ensure a fair and controlled comparison.

\subsection{Details of Method-Specific Setting}
\label{sec:specific_setting}
Beyond the general configuration described above, several methods require additional method-specific settings, which we describe in detail below.

\paragraph{IRCoT.}
IRCoT interleaves document retrieval with chain-of-thought reasoning. We employ the one-shot demonstrations provided in the original work, and set the maximum number of reasoning--retrieval iterations to 2.\footnote{\url{https://github.com/StonyBrookNLP/ircot/blob/main/prompts/2wikimultihopqa/gold_with_3_distractors_context_cot_qa_codex.txt}}

\paragraph{AAR.}
For the Augmentation-Adapted Retriever (AAR), we directly use the pre-trained retriever checkpoints released by the authors.\footnote{\url{https://huggingface.co/OpenMatch/AAR-Contriever-KILT}} Specifically, we adopt the AAR-Contriever-KILT variant for all experiments, without performing any additional fine-tuning.

\subsection{Details of Retrieval Quality Scoring Prompt}
\label{sec:retrieval_quality_prompt}

To supervise retrieval quality independently of downstream generation, we employ a LLM as an automatic evaluator to assess the quality of retrieved documents.
Given an input question and a set of retrieved documents, the evaluator is prompted to score the overall retrieval quality along multiple dimensions.

\paragraph{Prompt Example.}
Below we provide an example of the prompt used for retrieval quality scoring.
The evaluator is instructed to act as a professional information retrieval
expert and to assess the retrieved documents from several complementary
perspectives, including relevance, coverage, accuracy, ranking quality,
and redundancy.

\begin{quote}
\small
\texttt{You are a professional information retrieval quality evaluation expert.
Please comprehensively evaluate the quality of documents returned by a retrieval
system for the following question.}

\vspace{0.5em}
\texttt{Question: \{question\}}

\vspace{0.5em}
\texttt{Retrieved Documents:}

\texttt{Document 1 (ID: \{id\}): \{content\}}

\texttt{Document 2 (ID: \{id\}): \{content\}}

\texttt{$\cdots$}

\vspace{0.5em}
\texttt{Please evaluate the overall retrieval quality from the following
dimensions:}

\texttt{(1) Relevance, (2) Coverage, (3) Accuracy, (4) Ranking Quality, and
(5) Redundancy.}

\vspace{0.5em}
\texttt{Output the evaluation strictly in the following JSON format:}

\texttt{\{}

\texttt{\ \ "relevance\_score": <1--5>,}

\texttt{\ \ "coverage\_score": <1--5>,}

\texttt{\ \ "accuracy\_score": <1--5>,}

\texttt{\ \ "ranking\_score": <1--5>,}

\texttt{\ \ "redundancy\_score": <1--5>,}

\texttt{\ \ "overall\_score": <average score>,}

\texttt{\}}

\texttt{Only output the JSON result and do not include any additional text.}
\end{quote}

The overall retrieval quality score is computed as the average of the five
dimension-level scores and is subsequently used to construct positive and
negative samples for contrastive training of the Retrieval Quality Encoder.
This design allows us to obtain fine-grained and stable supervision signals
that are decoupled from generation errors.

\subsection{Details of Implementation} \label{sec:implementation_details}
For \ours{} architecture, we use \texttt{qwen3-embedding-0.6b}~\citep{DBLP:journals/corr/abs-2506-05176} as both the Retrieval Quality Encoder $\phi_r$ and the Generation Capability Encoder $\phi_g$. Encoders of the same type share parameters across queries and retrievers to ensure a consistent embedding space. Both the retrieval quality and generation utility embeddings have a dimension of 1024. To mitigate overfitting, we freeze all parameters of $\phi_r$ and $\phi_g$ except for the word embedding layer and the last two transformer layers during training. The encoders are optimized using AdamW~\citep{DBLP:conf/iclr/LoshchilovH19} with a learning rate of $2\times10^{-6}$ and an effective batch size of 512 (16 per device $\times$ 4 gradient accumulation steps $\times$ 8 GPUs), and are trained for 10 epochs. Training employs the InfoNCE loss with DeepSpeed ZeRO-3 optimization. We split the dataset with a validation ratio of 0.05. All experiments are conducted locally on 8 NVIDIA A100 GPUs.

\begin{table*}[t]
\centering
\small
\caption{Throughput and approximate cost per 10k supervision items (1$\times$8$\times$A100 node, vLLM; $\approx$\$5.68/hour). Qwen3-8B and LLaMA3-8B-Instruct are used for generation-utility supervision, while Qwen3-Next-80B-A3B-Instruct is used as the retrieval-quality judge. RQ and GU denote the Retrieval Quality and Generation Utility, respectively.}
\label{tab:supervision_cost}
\setlength{\tabcolsep}{3mm}
\begin{tabular}{l l c c c c}
\toprule
\textbf{Supervision Role} & \textbf{Model} & \textbf{items/s} & \textbf{Input toks/s} & \textbf{Output toks/s} & \textbf{Approx. cost / 10k} \\
\midrule
GU      & Qwen3-8B                    & 363.60 & 7,760.91  & 12,132.31 & $\approx$ \$0.04 \\
GU      & LLaMA3-8B-Instruct          & 523.62 & 26,336.37 & 4,119.47  & $\approx$ \$0.03 \\
RQ Judge & Qwen3-Next-80B-A3B-Instruct & 21.20  & 31,154.70 & 1,399.88  & $\approx$ \$0.74 \\
\bottomrule
\end{tabular}
\end{table*}

\subsection{Details of Supervision Throughput, Cost, and Reproducibility}
\label{app:supervision_cost}

To make the supervision overhead of \ours{} more transparent and reproducible, we report the measured throughput of each supervision component and provide an approximate cost per 10k supervision items. All measurements are conducted on one 8$\times$A100 node with vLLM acceleration, whose hourly cost is approximately \$5.68 in our environment.

For generation-utility supervision, one item corresponds to generating one RAG answer for a $(q, R_i)$ pair, from which EM/F1-based utility labels are derived. For retrieval-quality supervision, one item corresponds to evaluating the retrieved documents for a query using the external LLM judge described in Appendix~A.4. The approximate cost per 10k items is computed as:
\[
\text{Cost per 10k items} = \frac{5.68}{3600} \times \frac{10000}{\text{items/s}}.
\]

Table~\ref{tab:supervision_cost} shows that generation-utility supervision is relatively inexpensive under our setup, while retrieval-quality supervision is more costly due to the stronger external evaluator. We emphasize that the judge is used only to construct the retrieval-quality signal and is intentionally decoupled from the end-to-end generation-utility supervision. This makes it straightforward to swap the evaluator model or prompt in future reproductions and extensions.

\section{Retriever Encoding Strategy}
\label{app:retriever-encoding}

In \ours{}, retrievers are represented as explicit, learnable entities within the capability encoding space, rather than being modeled through textual metadata or sampled retrieval outputs.

\paragraph{Retriever Representation.}
Each retriever $R_i \in \mathcal{R}$ is assigned a unique special token $\langle \texttt{RET}_i \rangle$, which serves as its sole identifier. This token is added to the tokenizer vocabulary with a trainable embedding. The retriever is thus represented as a single-token input:
\[
x_{R_i} = [\langle \texttt{RET}_i \rangle].
\]
No retriever descriptions, corpus statistics, or retrieval traces are used.

\paragraph{Encoding and Parameter Sharing.}
Both queries and retrievers are encoded by the same capability encoders. Specifically, the Retrieval Quality Encoder $\phi_r$ and the Generation Utility Encoder $\phi_g$ are shared across all queries and retrievers:
\[
\mathbf{r}_q = \phi_r(q), \quad \mathbf{g}_q = \phi_g(q),
\]
\[
\mathbf{r}_{R_i} = \phi_r(\langle \texttt{RET}_i \rangle), \quad \mathbf{g}_{R_i} = \phi_g(\langle \texttt{RET}_i \rangle).
\]
The only retriever-specific parameters are the embeddings of the retriever tokens, ensuring that queries and retrievers lie in a unified metric space.

\paragraph{Discussion.}
This design models retrievers as learned latent capability embeddings, optimized end-to-end via contrastive supervision. It enables efficient encoding, avoids reliance on heuristic retriever features, and allows new retrievers to be incorporated by simply introducing new tokens without modifying the architecture.

\begin{table}[t]
\centering
\small
\setlength{\tabcolsep}{2pt}
\renewcommand{\arraystretch}{1}
\begin{tabular}{cc c c c c c c c c}
\toprule
\multirow{2}{*}{$\boldsymbol{\beta}$}
& \multirow{2}{*}{$\boldsymbol{\gamma}$}
& \multicolumn{2}{c}{\textbf{TQA}}
& \multicolumn{2}{c}{\textbf{NQ}}
& \multicolumn{2}{c}{\textbf{HQA}}
& \multirow{2}{*}{$\boldsymbol{\Delta}$\textbf{EM}}
& \multirow{2}{*}{$\boldsymbol{\Delta}$\textbf{F1}} \\
\cmidrule(lr){3-4}\cmidrule(lr){5-6}\cmidrule(lr){7-8}
 &  & \textbf{EM} & \textbf{F1}
 & \textbf{EM} & \textbf{F1}
 & \textbf{EM} & \textbf{F1}
 &  &  \\
\midrule
\textbf{0.2} & \textbf{0.2} & \textbf{62.24} & \textbf{71.52} & \textbf{37.41} & \textbf{48.72} & \textbf{29.05} & \textbf{41.33} & \textbf{0.00} & \textbf{0.00} \\

0.2 & 0.0 & 61.66 & 70.18 & 36.81 & 47.39 & 28.44 & 39.99 & -0.59 & -1.34 \\
0.2 & 0.1 & 61.94 & 70.99 & 37.14 & 48.21 & 28.78 & 40.82 & -0.28 & -0.52 \\
0.2 & 0.3 & 62.13 & 71.37 & 37.29 & 48.56 & 28.94 & 41.16 & -0.11 & -0.16 \\
0.2 & 0.4 & 61.88 & 70.98 & 37.06 & 48.19 & 28.69 & 40.83 & -0.36 & -0.52 \\
0.2 & 0.5 & 61.12 & 70.19 & 36.27 & 47.41 & 27.91 & 40.03 & -1.13 & -1.32 \\
\midrule
0.0 & 0.0 & 59.44 & 68.41 & 34.62 & 45.63 & 26.27 & 38.22 & -2.79 & -3.10 \\
0.0 & 0.1 & 61.13 & 69.88 & 36.28 & 47.09 & 27.94 & 39.71 & -1.12 & -1.63 \\
0.0 & 0.2 & 61.54 & 70.28 & 36.73 & 47.46 & 28.36 & 40.07 & -0.69 & -1.25 \\
0.0 & 0.3 & 61.22 & 70.01 & 36.37 & 47.21 & 28.02 & 39.81 & -1.03 & -1.51 \\
0.0 & 0.4 & 60.98 & 69.79 & 36.14 & 46.98 & 27.78 & 39.63 & -1.27 & -1.72 \\
0.0 & 0.5 & 60.57 & 69.64 & 35.74 & 46.84 & 27.39 & 39.47 & -1.67 & -1.87 \\
\midrule
0.4 & 0.0 & 61.92 & 70.83 & 37.07 & 48.04 & 28.72 & 40.66 & -0.33 & -0.68 \\
0.4 & 0.1 & 62.01 & 70.97 & 37.18 & 48.19 & 28.82 & 40.78 & -0.23 & -0.54 \\
0.4 & 0.2 & 62.04 & 71.07 & 37.19 & 48.27 & 28.84 & 40.86 & -0.21 & -0.46 \\
0.4 & 0.3 & 62.09 & 71.21 & 37.26 & 48.43 & 28.89 & 41.04 & -0.15 & -0.30 \\
0.4 & 0.4 & 61.78 & 71.02 & 36.96 & 48.24 & 28.61 & 40.84 & -0.45 & -0.49 \\
0.4 & 0.5 & 61.41 & 70.56 & 36.57 & 47.74 & 28.23 & 40.36 & -0.83 & -0.97 \\

\bottomrule
\end{tabular}
\caption{Sensitivity analysis of Generation Utility Encoder weights. }
\label{tab:gu_weight_sensitivity_data}
\end{table}

\begin{table}[t]
\centering
\small
\caption{End to end latency of different routing methods, where Lat. donetes Latency.}
\label{tab:routing_methods_latency}
\setlength{\tabcolsep}{2pt}
\begin{tabular}{lccc}
\toprule
\multirow{2}{*}{\textbf{Methods}} & \textbf{Min} & \textbf{Avg} & \textbf{Max} \\
 & \textbf{Lat. (ms)} & \textbf{Lat. (ms)} & \textbf{Lat. (ms)} \\
\midrule
\ours{} & 7.51 & 28.37 & 119.34 \\
Random  & 7.23 & 38.79 & 102.66  \\
Oracle Single Best & 17.81 & 18.54 & 18.89 \\

\bottomrule
\end{tabular}
\end{table}

\begin{table*}[t]
\centering
\small
\setlength{\tabcolsep}{6pt}
\begin{tabular}{p{1.5cm} p{2.5cm} p{1.0cm} p{2.5cm} p{2.6cm} p{4.0cm}}
\toprule
\textbf{Query} & \textbf{Question} & \textbf{Gold} &
\textbf{Non-RAG} & \textbf{RAG (selected)} & \textbf{Retrieved evidence (selected)} \\
\midrule
\texttt{test\_435} &
Which English word comes from an old French word meaning sour wine? &
Vinegar &
Vinegar (EM=1.0) &
Qwen3-emb-0.6B $\to$ Sack;\;
E5 $\to$ Bastardo;\;
BM25 $\to$ Cheese &
\textbf{Distracting:} ``\emph{Sack (wine)} ... derived from French \emph{sec} (dry)'';\;
``\emph{Trousseau} ... also known as \emph{Bastardo}'';\;
``\emph{Cheese} ... root meaning `become sour' '' \\
\midrule
\texttt{test\_9847} &
Which perennial herb appeared on the twelve sided three penny bit coin? &
Thrift &
Thistle (EM=0.0) &
Qwen3-emb-4B $\to$ Thrift plant (Acc=1.0);\;
DIVER $\to$ Thrift plant (Acc=1.0);\;
BM25 $\to$ Loonie &
\textbf{Helpful:} ``reverse shows a \emph{three-headed thrift plant}'' (History of the threepence);\;
\textbf{Off-topic:} Canadian penny / loonie pages; unrelated plants (e.g., \emph{Protea caffra}) \\
\bottomrule
\end{tabular}
\caption{Case study with retrieved evidence summaries. For readability, we show representative retrievers and one discriminative snippet per outcome (helpful vs.\ distracting).}
\end{table*}

\section{Extended Analysis}

\subsection{Extended Analysis of Parameters}\label{sec:details_of_parameters}
This subsection reports the complete numerical results for the sensitivity analysis of the generation utility weighting coefficients $\beta$ and $\gamma$ in Eq.~(5), as summarized in Table~\ref{tab:gu_weight_sensitivity_data}. We evaluate different combinations of $\beta$ and $\gamma$ on three downstream benchmarks (TQA, NQ, and HQA), and report both EM and F1 scores, along with their absolute differences relative to the default setting $\beta=\gamma=0.2$. The results show that removing either query-level answer supervision ($\beta=0$) or retriever-level global supervision ($\gamma=0$) consistently degrades performance, while removing both leads to a substantial drop across all datasets. Introducing non-zero values for either coefficient partially recovers performance, but the best and most stable results are achieved when both signals are combined with moderate weights. In particular, settings with $\beta$ and $\gamma$ in the range of $0.1$ to $0.3$ yield comparable performance, whereas overly small or large values result in suboptimal outcomes. These observations indicate that query-level answer quality and retriever-level global correctness provide complementary but non-redundant supervision signals for generation utility modeling. Based on this analysis, we adopt $\beta=\gamma=0.2$ in all experiments.

\subsection{Extended Analysis of Efficiency}\label{sec:extended_analysis_of_efficiency}

We further analyze the end-to-end efficiency of different routing strategies, with results reported in Table~\ref{tab:routing_methods_latency}. The latency is measured in an end-to-end manner, including both the routing overhead (retriever selection and retrieval) and the downstream generation time. For all methods, the generation component is executed locally using the \texttt{vLLM} framework~\cite{DBLP:conf/sosp/KwonLZ0ZY0ZS23} with identical decoding configurations, ensuring a fair comparison under a high-throughput inference setting.

As shown in Table~\ref{tab:routing_methods_latency}, the proposed method achieves competitive average latency while maintaining a low minimum latency comparable to Random routing, indicating negligible routing overhead in favorable cases. Although the maximum latency of our method is higher due to dynamic retriever selection and varying generation lengths, it remains within a reasonable range and does not significantly impact overall efficiency. In contrast, Random routing exhibits higher average latency, reflecting its lack of informed selection, while Oracle Single Best shows stable but higher minimum latency, as it consistently routes all queries to a fixed retriever regardless of query difficulty. Overall, these results demonstrate that our approach strikes a favorable balance between routing adaptivity and end-to-end efficiency.

\section{Case Study}
\label{sec:case_study}
We analyze two representative TriviaQA cases to highlight (i) scenarios where retrieval is unnecessary and even detrimental, and (ii) scenarios where retrieval is essential but highly sensitive to retriever quality. Together, these cases motivate query-adaptive routing and retriever-aware decision making.

\paragraph{Case 1: Retrieval Can Be Unnecessary and Harmful.}
For some queries, the generator’s parametric knowledge is sufficient to produce the correct answer. Introducing retrieval in such cases may inject misleading evidence and shift the generation distribution. As a result, RAG can underperform non-RAG baselines, indicating that routing mechanisms should allow skipping retrieval when confidence in parametric knowledge is high.

\paragraph{Case 2: Retrieval Is Necessary but Retriever Capability Matters.}
Other queries fundamentally depend on external evidence. In these cases, effective retrievers surface precise and discriminative information that enables correct generation, while weaker retrievers return noisy or irrelevant documents that still lead to failure. This underscores that, beyond deciding whether to retrieve, routing must account for retriever-specific capability.

\section{Usage of AI Assistants}
We use ChatGPT to improve the presentations of this paper.\footnote{\url{https://chatgpt.com/}}

\end{document}